\documentstyle[12pt,titlepage]{article}
\begin{document}
\begin{center}
{\bf The Painlev\'e Integrability Test}
\vskip 5pt
Willy Hereman \\
Department of Mathematical and Computer Sciences, \\
Colorado School of Mines, Golden, CO 80401-1887, U.S.A.
\end{center}
\vskip 5pt
The Painlev\'e test is a widely applied and quite successful technique 
to investigate the integrability \cite{flanewtab} of nonlinear ODEs and PDEs
by analyzing the singularity structure of the solutions. 
The test is named after the French mathematician Paul Painlev\'e 
(1863-1933) \cite{levwin}, who classified second order differential 
equations that are solvable in terms of known elementary functions 
or new transcendental functions \cite{ince}.  

The Painlev\'e test, allows one to verify whether or not a
differential equation (perhaps after a change of variables)
satisfies the necessary conditions for having the Painlev\'e property.
If so, the equation is prime candidate for being completely integrable 
\cite{ablcla}.

As originally formulated by Ablowitz {\it et al.} 
\cite{ablramseg}, the Painlev\'e conjecture asserts 
that all similarity reductions of a completely integrable PDE 
should have the Painlev\'e property (or be of Painlev\'e-type),
i.e.\ their general solutions should have no movable singularities 
other than poles in the complex plane. 

A later version of the Painlev\'e test due to Weiss {\it et al.} 
\cite{weitabcar} allows testing of PDEs directly, without recourse 
to the reduction(s) to ODEs.
A PDE is said to have the Painlev\'e property if its 
solutions in the complex plane are single-valued in the neighborhood of 
all its movable singularities. In other words, the equation must have
a solution without any branching around the singular points whose
positions depend on the initial conditions.
The traditional Painlev\'e test does not test for essential singularities
and therefore cannot determine whether or not branching occurs about these. 
\vskip 5pt
\noindent
{\bf The algorithm}
\vskip 2pt
\noindent
The Painlev\'e test can be applied to nonlinear polynomial system of 
ODEs or PDEs with (real) polynomial terms. For brevity, we give the 
three steps of the test for a single PDE, ${\cal F} (x,t,u(x,t)) = 0,$ 
in two independent variables $x$ and $t$. 

Following \cite{weitabcar}, the Laurent expansion of the solution $u(x,t)$, 
\begin{equation} 
\label{pdelaurent}
u(x,t) = g^{\alpha}(x,t) \sum_{k=0}^{\infty} u_{k}(x,t) \; g^{k}(x,t), 
\end{equation} 
should be single-valued in the neighborhood of a non-characteristic, 
movable singular manifold $g(x,t), $ which can be viewed as the 
surface of the movable poles in the complex plane.
In (\ref{pdelaurent}), $u_{0} (x,t) \ne 0$, $\alpha$ is a negative integer,
and $u_{k}(x,t) $ are analytic functions in a neighborhood of $g(x,t).$

Note that for ODEs the singular manifold is $g(x,t) = x - x_0,$ 
where $x_0$ is the initial value for $x$.
For PDEs, if $u(x,t)$ has simple zeros and $g_{x}(x,t) \ne 0,$ 
one may apply the implicit function theorem near the singularity manifold 
and set $g(x,t) = x - h(t),$ for an arbitrary function $h(t)$ 
\cite{kruramgra,ramgrabou}. This considerably simplifies the computations.
\vfill
\newpage
\noindent
{\it Step 1: Leading order analysis}
\vskip 3pt
\noindent
Determine the (negative) integer $\alpha$ and $u_0$ by balancing the 
minimal power terms after substitution of $u = u_0 g^{\alpha}$ into 
the given PDE.
There may be several branches for $u_0,$ and for each the next two steps 
must be performed. 
\vskip 5pt
\noindent
{\it Step 2: Determination of the resonances}
\vskip 2pt
\noindent
For a selected $\alpha$ and $u_0,$ calculate the non-negative integers $r$, 
called the {\it resonances}, at which arbitrary functions $u_r$ enter 
the series (\ref{pdelaurent}). 
To do so, substitute $u = u_0 g^{\alpha} + u_r g^{\alpha+r}$ into the 
equation, only retaining its most singular terms.
Require that the coefficient $u_r$ is arbitrary by equating its 
coefficient to zero. 
Compute the integer roots of the resulting polynomial.
For (\ref{pdelaurent}) to represent the general solution, the number of roots
(including $r=-1)$ must match the order of the given equation. 
The root $r=-1$ corresponds to the arbitrariness of the manifold $g(x,t).$
\vskip 5pt
\noindent
{\it Step 3: Verification of the compatibility conditions}
\vskip 2pt
\noindent
Verify that a solution of the form (\ref{pdelaurent}) is indeed admissible, 
and that it has the necessary number of free coefficients $u_r.$ 
Substitute (\ref{pdelaurent}), truncated a the largest resonance, 
into the PDE. Determine $u_k$ at non-resonance levels $k.$
At resonance levels, $u_r$ should be arbitrary, and since we are 
dealing with a nonlinear equation, a {\it compatibility condition} must be 
unconditionally satisfied. 

An equation for which these three steps can be carried out consistently and 
unambiguously passes the Painlev\'e test.
\vskip 5pt
\noindent
In the case of systems, for every dependent variable $u_i$ one substitutes  
\begin{equation}
\label{odelaurent}
u_i = g(x,t)^{\alpha_i} \sum_{k=0}^{\infty} u_k^{(i)} g(x,t)^k, 
\end{equation}
and carefully determines all branches of dominant behavior corresponding 
to various choices of $\alpha_i$ and/or $u_0^{(i)}.$ 
For each branch, the single-valuedness of the corresponding Laurent 
expansion must be tested, i.e. the resonances must be computed and the 
compatibility conditions must be verified. 
Details and an abundance of  worked examples can be found in 
\cite{ablcla,conte,contebook,flanewtab,kruramgra,ramgrabou,steeul}. 
\vskip 5pt
\noindent
{\bf Simple Examples}
\vskip 3pt
\noindent
Consider the PDE, $ u_{tx} + a(t) u_x + 6 u u_{xx} + 6 u_x^2 + u_{xxxx} = 0,$ 
and ask under what condition for $a(t)$ the equation passes the Painlev\'e 
test.

Here, $\alpha = -2$ and $u_0 = - 2 g_x^2.$ 
Apart from $r=-1,$ the roots are $r=4, 5,$ and $6.$
The latter three are resonances. Furthermore, $u_1, u_2$ and $u_3$ can
uniquely be determined in terms of derivatives of $g(x,t).$

The compatibility conditions at resonances $r=4$ and $r=5$ are 
satisfied. Hence, $u_4$ and $u_5$ are arbitrary.  
The compatibility condition at resonance $r=6$ is $ a_t + 2 a^2 = 0. $
Hence, $a = \frac{1}{2t}$ and the PDE becomes the cylindrical KdV 
equation which is indeed completely integrable \cite{ablcla}.

As a second example, consider the famous Lorenz system from meteorology,
\begin{equation}
\label{lorenz}
u_1' = a (u_2 - u_1), 
\;\quad 
u_2' = - u_1 u_3 + b u_1 - u_2, 
\;\quad 
u_3' = u_1 u_2 - c u_3,
\end{equation}
where $a,b,$ and $c$ are positive constants. 

For each dependent variable, one substitutes a Laurent series 
(\ref{odelaurent}) and determines the leading orders: 
$\alpha_1 = -1, \alpha_2 = \alpha_3 = -2.$
The first coefficients are $u_0^{(1)}= \pm 2 i, 
u_0^{(2)}=\mp 2 i/a, u_0^{(3)} = -2/a. $
The roots are $r=-1, 2, 4.$ The expressions for 
$u_1^{(1)}, u_1^{(2)}$ and $u_1^{(3)}$ are readily computed.

The compatibility conditions at resonances $r=2$ and $r=4$ are not
satisfied. At resonance $r=2$ one encounters $a (c - 2a) (c + 3a - 1) = 0.$ 
Investigating all cases, it turns out that for $c=2a$ 
the compatibility condition at $r=4$ is not satisfied.
For $c=1-3a,$ the compatibility condition at $r=4$ is satisfied if
$a = \frac{1}{3}.$
The Lorenz system (\ref{lorenz}) thus passes the Painlev\'e test 
when $a=\frac{1}{3}$ and $c=0$ \cite{heretal}.  

In the last example, we consider a coupled system of KdV equations, 
\begin{equation}
\label{hirsat}
u_{1,t} - 6 a u_1 u_{1,x} + 6 u_2 u_{2,x} - a u_{1,xxx} = 0, \;\quad 
u_{2,t} + 3 u_1 u_{2,x} + u_{2,xxx} = 0,
\end{equation}
where $a$ is a nonzero parameter. 
System (\ref{hirsat}) is known to be completely integrable 
if $a=\frac{1}{2}.$ This is confirmed by the Painlev\'e test. 
Indeed, with a Laurent series for $u_1$ and $u_2$ one obtains 
$\alpha_1 = \alpha_2 = -2$ and $r=-2,-1,3,4,6$ and $8.$
Furthermore, $u_0^{(1)}=-4$ and $u_0^{(2)}=\pm 2 \sqrt{2 a} $ determine the 
coefficients $u_1^{(1)}, u_1^{(2)}, u_2^{(1)}, u_2^{(2)}$ unambiguously. 
At resonances $3$ and $4$ there is one free function and no condition 
for $a.$ The coefficients $u_5^{(1)}$ and $u_5^{(2)}$ are unique determined.
At resonance $6,$ the compatibility condition is only satisfied if 
$a = \frac{1}{2}.$ For this value, the compatibility condition at $r=8$ 
is also satisfied. 
\vskip 5pt
\noindent
{\bf Symbolic Programs}
\vskip 2pt
\noindent
The Painlev\'e test, although algorithmic, is cumbersome when done by hand. 
Several computer implementations of the Painlev\'e test exist 
\cite{conte,heretal,herzhu}. A brief review is given in \cite{scheen}.  
These symbolic codes are particularly useful for the verification of 
the self-consistency (compatibility) conditions, and in exploring 
all possibilities of balancing singular terms. 
Applied to equations with parameters, the software can determine the 
conditions on the parameters so that the equations pass the Painlev\'e 
test (see \cite{heretal,herzhu}).
\vskip 5pt
\noindent
{\bf Further Reading}
\vskip 3pt
\noindent
There is a vast amount of literature about the test and its applications 
to specific differential equations. Several well-documented surveys 
\cite{cartab,conte,flanewtab,gibetal,krucla,krujoshal,laksah,newtabzen}
and books \cite{clarksonbook,contebook,steeul} discuss the basics, as well
as subtleties and pathological cases of the test. 
The survey papers also deal with the many interesting connections 
with other properties of PDEs and by-products of the Painlev\'e test.
They show, for example, how truncated Laurent series expansions 
allow one to construct Lax pairs, B\"acklund and Darboux transformations, 
and closed-form particular solutions of PDEs.

Some shortcomings of the traditional Painlev\'e test have been identified 
by Kruskal and others \cite{kru,krucla,krujoshal}. 
Improved versions of the Painlev\'e test have been proposed, such as 
the poly-Painlev\'e test \cite{krucla}. 
Besides, other variants of the test exist 
\cite{conte,contebook,kru,krujoshal}, e.g the weak Painlev\'e test 
\cite{ramgrabou}, and a perturbative Painlev\'e approach \cite{conforpic}
which allows for a deeper analysis of equations with negative resonances.

\end{document}